\documentclass[10pt,fleqn,oneside]{article}

  \newtheorem{theorem}{Theorem}
 
\newtheorem{nn}[theorem]{$\!$\hspace{-2pt}}
\newtheorem{apnn}[theorem]{{\bf A\hspace{-2pt}}}

  \newenvironment{proof}{{\it Proof.\ \/}}{\  }

  \newcommand{\bb}{\begin{nn}\em }
\newcommand{\ee}{\end{nn}}
  
   \newcommand{\apbb}{\begin{apnn}\em }
\newcommand{\apee}{\end{apnn}}

\renewcommand{\a}{\`{a}\ }

 \newcommand{\up}{\uparrow}

      \newcommand{\Ra}{\Rightarrow}

 \newcommand{\ap}{'}

 \newcommand{\qm}[1]{\mbox{``${#1}$''}}

\begin{document}

 \thispagestyle{empty}
 \pagestyle{plain}
 \title{{\Large Kleene, Rogers and Rice Theorems\\ Revisited in C and in Bash}}
  
   \author{
   Salvatore Caporaso\\
{\small  (caporaso@di.uniba.it)}\\
  Nicola Corriero\\
{\small  (nicolacorriero@gmail.com)}
\\
     {\small Dipartimento d'Informatica
dell\ap Universit\a di Bari}
    }

   \date{}
 
  \maketitle
\begin{abstract}
The recursion theorem in the weak form $\{e\}(z)=x(e,z)$ (universal function not needed)
and in Rogers form $\phi_{\phi_{f(n)}}(z)=\phi_n(z)$, and Rice theorem are proved
a first time using
programs in C, and a second time with scripts in Bash.
\end{abstract} 
\section{Introduction}
One of the cornerstones of recursion theory is
the result known as $S-m-n$ theorem (in honour of the original notation
by Kleene, who  called it \textit{Iteration Theorem})
or as \textit{Parameter Theorem} (after Schoenfield). Its
proofs  in  Literature however are not fully
satisfatory for a computer scientist. Some authors
merely appeal to Church thesis (Rogers \cite{Rog}, Cutland \cite{Cut}, Enderton \cite{End}).
Some others arithmetize the metaprocessing, and this  disguise the  computation under
a misleading  plenty of numerical technicalities 
(Kleene \cite{IMM}, Smorynski\cite{Smo}). The proof by Kechris and Moschovakis
\cite{KM} use a universal function, which is not available  when
classes of total functions are discussed,  like in the case, for example, of security or complexity classes.
Among the consequences of the $S-m-n$
we have the Kleene weak form of the Recursion Theorem (existence of a fixed-point value),
the Rogers form (functional fixed-point), Rice theorem, 
the analysis   by Thompson \cite{Thom}
in his Turing  lecture of relationship  between malware 
and Quine's indirect self-referential paradox.
We feel that simple idea need simple programs, and that, therefore,  understanding these phenomena needs their revisitation
in terms of \textit{real programming}. In this paper we show that this is rather
easy in C; and even easier in a language allowing quick writing of rough programs
like Bash.
To this purpose, we  code, in both these languages, the procedures needed
to prove the results mentioned above.

A by-result of this work is that one can show significant results in a couple of lecures in the context of a beginners
programming course: there is no need of the cumbrous paraphernalia of abstract models of computation like
TMs, recursive functions or functional programming.

Familiaritity with C and/or Bash is not needed to follow the broad outlines of our discussion.
To check the details one needs the small amount of information which is contained in
  Kernighan \& Ritchie \cite[Ch. 1  \textit{A Tutorial Introduction} pp.5-30]{KR}.
All additional details about C,
and the essential parts for Bash are explained by means of examples.

\section{In C}
	 	
\bb\textbf{Notation} \ (1) $\Sigma$ is the set of all strings
in the alphabet of all characters that may occur in a C source file.  $\phi^{(n)}$   is a (partial) $n$-ary
function such that 
$\phi^{(n)}:\Sigma^n\mapsto\Sigma$
($n=1,2$ often omitted). $\mathtt{\phi(y)=u}$
is short for $\mathtt{(y,z)\in\phi}$ or $\phi(\mathtt{y})\up$.

(2)\quad  \texttt{a, b} and \texttt{c} are autonymous names for the three fixed identifiers
consisting of resp.  the 1st, 2d and 3d low-case letter of the Latin alphabet.
They will play a crucial role throughout this paper.

(3)\quad  \texttt{r,\ldots,z}, possibly followed by decimal digits, are generic or constant
strings.

(4)\quad We are going to discuss the behaviour of
certain (C functions defined by) strings \texttt{x}
of the form
\begin{equation}
\label{fn_}
\mathtt{\mathtt{x\_}()\{\ldots}
\end{equation}
where \texttt{x\_} denotes the identifier  used in calls to \texttt{x}.
To this purpose, we write
 \texttt{a=y} to  mean that the string variable \texttt{a} is assigned with \texttt{y}.
And we write  \texttt{x:y,z=u}  if after a call \texttt{x\_();}
with \texttt{a=y} and with \texttt{b=z} we get \texttt{c=u}.
Calls are tacitly assumed to be syntactically correct, and to include all needed directives and declarations.
\ee

\bb\label{phi}\textbf{Definition}\quad
 String \texttt{x} of the form (\ref{fn_})
 \textit{standard computes}
(\textit{s-computes})  function $\phi$  if  we have
\[
\mathtt{x:y,z=u}\quad \quad \mbox{iff}\quad \quad \mathtt{u=\phi(y,z)}
\]
  (\texttt{z}  absent, and \texttt{b} immaterial for $n=1$).

\textbf{Notation}\quad  $\mathtt{\phi_x}$ is the function s-computed by \texttt{x}.

\textbf{Example}\quad Let \texttt{id} and \texttt{s1} denote resp. the strings
(see \S\ref{string.h} for \texttt{strcpy})
\begin{verbatim}
id_(){                           
  strcpy (c,a);                    
}             
s1_(){
  strcpy(c,a);
  strcpy(b,a);  // comment: just to use two variables
}                    
\end{verbatim}
 we have \texttt{id:y=y} and \texttt{s1:y,z=y} and, therefore,
\[
\mathtt{\phi_{id}(y)=y\quad \quad \quad \phi_{s1}(y,z)=y}
\]
\ee\bb\label{string.h}\textbf{Summary of string functions from the  standard library}
Recall that the following functions are defined in $\mathtt{<string.h>}$
\begin{verbatim}
     strcpy(s,t)                       s = t     (i.e. t is copied into s)
     strcat(s,t)                       s = st    (concatenation)
     strchr(s,'c')                     locates the 1st occurrence in s of character 'c'
     strncpy(s,t,i)                    s = first i characters of t (doesn't add '\0')
\end{verbatim}
In what follows we need the C function defined by \texttt{fn=}
\begin{verbatim}
fn_(){
  int i;
  i=strchr(s,'()');    // pointer to the leftmost parenthesis
  strncpy(t,s,i);
  t[i]='\0';          // because strncpy doesn't do it
}
\end{verbatim}
\texttt{fn} takes the definition of a function into the fuction name, in the sense that we have
\begin{verbatim}
     fn:x_(){... = x_
\end{verbatim}
\ee
\bb\label{smn}
\textbf{Diagonal Substitution Lemma}\quad (A variant of the \textit{s-m-n} theorem).
There is a C function \texttt{ds} which s-computes the function $\mathtt{\sigma^{(1)}}$ such that for all $\mathtt{\phi_x^{(2)}(y,z)}$  we have
\[
\mathtt{\phi_{\sigma(x)}(y)=\phi_x(x,y)};\quad \mbox{or, in other terms,}\quad
\mathtt{ds:x=u}\quad \mbox{implies}\quad \mathtt{\phi_u(y)=\phi_x(x,y)}
\]
\ee
\proof
Define \texttt{ds=}  \begin{verbatim}	
ds_(){
  fn_();                                      // c=x_
  strcpy(b,c);                               // b=x_ 
  strcpy(c,"s_(){strcpy(b,a);strcpy(a,\"");  // c=s_(){strcpy(b,a);strcpy(a,"
  strcat(c,a);                               // c=s_(){strcpy(b,a);strcpy(a,"x
  strcat(c,"\");");                          // c=s_(){strcpy(b,a);strcpy(a,"x");
  strcat(c,b);                       // c=s_(){strcpy(b,a);strcpy(a,"x");x_
  strcat(c,"();}");                  // c=s_(){strcpy(b,a);strcpy(a,"x");x_();}
  strcat(c,a);                       // c=s_(){strcpy(b,a);strcpy(a,"x");x_();}x
}
\end{verbatim}
The comments above (at the left of the \verb+//+'s) show that we have 
 \texttt{ds:x=u} with (up to unnecessary new lines and indentations  added for the sake of
 readability) 
\begin{verbatim}
\texttt{s=}
s_(){
  strcpy(b,a);
  strcpy(a,"x");
  x_();
}
x                                         
\end{verbatim}  
So, for \texttt{a=y}, (1) we put \texttt{b=y} and \texttt{a=x}; and
(2) we call \texttt{x} with these new values.
Hence \texttt{x:x,y=w} implies, as promised, \texttt{u:y=w}.

\bb\label{weak}\textbf{Kleene Theorem}\quad  (A weak form of the Second  Kleene Theorem)
For each  $\mathtt{\phi^{(2)}_x(y,z)}$ there is a \textit{fixed point}
\texttt{u} such that
\[\mathtt{
\phi_u(z)=\phi_x(u,z).
}\]
\ee  
\proof
Given \texttt{x} in the form (\ref{fn_}), define a new  C function by the string
\texttt{x0=}
\begin{verbatim}
x0_(){
  ds_();
  strcpy(a,c);
  x_();
}
ds
x
\end{verbatim} 
We have
\begin{equation}
\label{kl1}
\mathtt{
\phi_{x0}(y,z)=\phi_x(\sigma(y),z)
}
\end{equation}
because (1) by calling \texttt{ds} with  \texttt{a=y} we get $\mathtt{c=\sigma(y)}$;
(2) by calling \texttt{x} with  $\mathtt{a=\sigma(y)}$
(via \texttt{strcpy(a,c)}) and with \texttt{b=z} we get $\mathtt{c=\phi_x(\sigma(y),z)}$.
Now define
\begin{equation}
\label{kl2}
\mathtt{
u=\sigma(x0)}
\end{equation}
The result follows because we have
\[
\mathtt{
\phi_u(z)=\phi_{\sigma(x0)}(z)=\phi_{x0}(x0,z)=\phi_x(\sigma(x0),z)=\phi_x(u,z)
}
\]
where we owe the first equality to definition (\ref{kl2}) and the second to
Lemma \ref{smn};  and where we get the last two from resp.
(\ref{kl1}), and   (\ref{kl2}) again.

\bb\textbf{Note}\quad
By applying the theorem to the \texttt{s1} of \S\ref{phi} we get a \textit{quine} in C,
i.e. a function definition that prints itself by means of
a so-called \textit{indirect self-reference} of the form
\[
\mbox{``\ldots'' what  quoted is \ldots}
\]
This quine includes a comment, which could be replaced by different actions of another kind.
\ee

\bb\label{univ}\textbf{The Universal Function}\quad
One can   write a string $\mathtt{univ=univ\_()\{\ldots}$   defining a C function
which s-computes a universal function, in the sense that we have
for all $\mathtt{\phi_x^{(1)}}$ and \texttt{y}
\begin{equation}
\label{eq_univ}
\mathtt{\phi_{univ}(x,y)=\phi_x(y).}
\end{equation}
The proof of Theorem \ref{weak} needs a few linear-time operations,
and, therefore, it holds for almost all total fragments of C.
We regard next theorem as a stronger
form of that theorem,  because its proof, being based on the existence of a universal function,
 fails with any class of total functions.
\ee

\bb\label{strong}\textbf{Rogers Theorem}\quad  (A strong form of the Second  Kleene Theorem)
For each   $\mathtt{\phi_x^{(1)}(y)}$
 there is a value \texttt{v} such that 
\[
\mathtt{\phi_{\phi_x(v)}=\phi_v}
\]
\ee
\proof Given \texttt{x} in the form (\ref{fn_}) define a new C function by means of the string \texttt{w=}
\begin{verbatim}
w_(){
  x_();
  strcpy(a,c);
  univ_();
}
univ
x
\end{verbatim} 
We have 
\begin{equation}\label{rg}
\mathtt{\phi_w(y,z)=\phi_{univ}(\phi_x(y),z)}
\end{equation}
because (1) by calling \texttt{x} with \texttt{a=y} we get
$\mathtt{c=\phi_x(y)}$; (2) by copying \texttt{c} into \texttt{a}
and calling \texttt{univ} with this value for \texttt{a} and with \texttt{b=z}
we obtain  (by (\ref{eq_univ}) with$\mathtt{\phi_x(y)}$ as \texttt{x})
\begin{equation}
\label{rg2}
\mathtt{c=\phi_w(y,z)=\phi_{\phi_x}(y)(z)}
\end{equation}
Our assertion follows by taking as \texttt{v} the fixed-poin for \texttt{w}
which is granted by Theorem \S\ref{weak}. Indeed, we then have
\[
\mathtt{
\phi_v(z)=\phi_w(v,z)=\phi_{univ}(\phi_x(v),z)=\phi_{\phi_x(v)}(z)
}
\]
where the first equality follows because \texttt{v} is the fixed-point for $\mathtt{\phi_w}$;
the second by
(\ref{rg}), and the last one by (\ref{rg2}).

%
\bb\textbf{Rice Theorem}\quad 
All not-trivial   classes of  s-computable functions are  undecidable.
\ee
\proof Assume (ad abs.) that there is a string \texttt{x=x\_()\{\ldots}
that s-computes the characteristic function of
${\cal A}$, in the sense that we have
\begin{equation}
\label{due+}
\mathtt{\phi_x(z)=\qm{0}\mbox{ iff }\phi_z\in{\cal A};\quad
\phi_y(z)=\qm{1}\mbox{ iff }\phi_z\not\in{\cal A}.}
\end{equation}
Since the class is not trivial
there exist \texttt{s, t} such that
\begin{equation}
\label{uno+}
\phi_\mathtt{s}\in{\cal A};\qquad\phi_\mathtt{t}\not\in{\cal A}.
\end{equation}
Define \texttt{y=}
\begin{verbatim}
y_(){
  x_();
  if strcmp(c,"0")
    then strcpy(c,"t");
  else strcpy(c,"s");
}
x
\end{verbatim} 
we have
\begin{equation}
\label{due+}
\mathtt{\phi_y(z)=t\mbox{ iff }\phi_z\in{\cal A};\quad
\phi_y(z)=s\mbox{ iff }\phi_z\not\in{\cal A}.}
\end{equation}
Let \texttt{u} be the string granted by Rogers Theorem, such that we have
\begin{equation}\label{rg3}
    \mathtt{\phi_u=\phi_{\phi_y(u)}}
\end{equation}
We get the following contradiction
 \[
 \begin{array}{llll}
 \phi_\mathtt{u}\in{\cal A}&\Ra&\mathtt{\phi_y(u)=t}&\mbox{equation (\ref{due+})}\\
&\Ra&\mathtt{\phi_u=\phi_t}&\mbox{equation (\ref{rg3})}\\
&\Ra&\mathtt{\phi_u\not\in{\cal A}}\qquad\qquad\qquad\qquad\qquad\qquad\qquad
&\mbox{equation (\ref{uno+})}\\
\mathtt{\phi_u\not\in{\cal A}}&\Ra&\mathtt{\phi_y(u)=s}&\mbox{equation (\ref{due+})}\\
&\Ra&\mathtt{\phi_u=\phi_s}&\mbox{equation (\ref{rg3})}\\
&\Ra&\mathtt{\phi_u}\in{\cal A}\qquad\qquad\qquad\qquad\qquad\qquad\qquad
&\mbox{equation (\ref{uno+})}.
 \end{array}
 \]

\section{In Bash}

\bb\label{bash_notat1}\textbf{Notation} \ (1) \ \texttt{foo()} is the string stored in file \texttt{foo}. When \texttt{foo()} is a script  \texttt{x}  we  display it along an indented column  (with semicolon omitted according to
 Bash syntax). For example,
\begin{verbatim}
eecho()=
  echo echo hi! > hi
  chmod 755 hi
  hi
\end{verbatim}  
says that file \texttt{eecho} contains a script
that: \textit{redirects} the output \texttt{echo hi!} of \texttt{echo echo hi!}
from \textit{stdout} (the monitor) to file \texttt{hi} ($\ell$1) ; grants the
execution permissions to  file \texttt{hi()=echo hi!} ($\ell$2); and calls it
($\ell$3).

(2)\quad We use the sign \verb+->+ to show the Bash \textit{prompt}. A line like
\begin{verbatim}
-> comm arg1 ... argk
\end{verbatim}
says that at the prompt command \texttt{comm}
 with  arguments \texttt{arg1 ... argk} ($k \geq 0$)
enters from \textit{stdin} (the console). Below such a line
we list the  $h\geq 0$ lines that the command sends to stdout and the  $k\geq 0$   created
files. The convention of part (1) allows the distinction between the former and the latter ones. For example, to   say that  \texttt{eecho} creates file \texttt{hi} and prints \texttt{hi!},
we write
\begin{verbatim}
-> eecho
hi()=
  echo hi!
hi!
\end{verbatim}
To summarize these notations:

 (\textit{i}) \quad  a not-indented column like
\begin{verbatim}
foo()
  x
\end{verbatim} 
 means that file \texttt{foo} stores \texttt{x};

(\textit{ii})\quad  the same column, below  \verb+->+\ldots says that \texttt{foo} has been
created by  the command line \ldots;

(\textit{iii})\quad \texttt{foo()} alone stands for \texttt{x};

(\textit{iv})\quad 
 a not-indented string  below a prompt  is an output.

(4)
\quad \texttt{comm1 args1  = comm2 args2}
says that
\texttt{comm1 args1} and
\texttt{comm2 args2} print the same string ---
differences in their other effects (f.i. in the created files)
do not matter.
 \ee

 \bb\label{bash_commands}\textbf{Summary of useful Bash commands} \   (1) \
Recall that Bash assigns
its internal variables \texttt{\$1, \$2,\ldots,\$n} with
the first, second,\ldots,$n$-th argument of the  script
being currently executed.
So, we have
\begin{verbatim}
id()=
  echo $1
-> id foo
foo
\end{verbatim} 
Since the command \texttt{cat foo bar}
sends to stdout the concatenation of  \texttt{foo()} and \texttt{bar()}, we have
\begin{verbatim}
cat2()=
  cat $1 $2
-> cat2 id cat2
echo $1
cat $1 $2
\end{verbatim}
(2)\quad 
Assume
\begin{verbatim}
-> comm args
u
\end{verbatim} 
Bash interprets an expression like \texttt{\$(comm args)} as a \textit{command substitution} of that same expression with \texttt{u}. For example
\begin{verbatim}
-> echo $(cat id)
echo $1
\end{verbatim} 
(3)\quad 
Command \texttt{set arg1 \ldots argk} assigns \texttt{arg1,\ldots,argk}
to \texttt{\$1,\ldots,\$k}. So we have
\begin{verbatim}
cat2idcat2()=
  set id cat2
  cat2
-> cat2idcat2
echo $1
cat $1 $2
\end{verbatim}
\ee

\bb\textbf{Note}\quad
In a script \texttt{builder} which produces another script \texttt{built}, we include
in \texttt{builder} the line \texttt{chmod 755 built}.
 In all other cases, we tacitly assume that the
execution permissions have been granted to the current script, when it has been edited.
\ee

 \bb
 \textbf{Scripts arity}\quad
The \textit{arity} of script  \texttt{x}
 is, by definition,  $n\geq 0$ if the variables
 \texttt{\$1,\ldots,\$n} occur in \texttt{x}.
So, the arity of the previously introduced scripts \texttt{cat2, id}
and \texttt{hi} is resp. 2, 1  and 0.
 \ee
\bb
\textbf{Notation}\quad $\mathtt{\varphi_x^{(n)}}$ ($n=1,2$)
is the function $\varphi:\Sigma^n\mapsto(\Sigma)$
such that we have
\begin{verbatim}
-> x y z
u
\end{verbatim}
iff $\mathtt{u=\varphi(y,z)}$ (\texttt{z} absent for $n=1$).

Note that we don't need any standard of computation now.
\ee
\bb \label{weak_bash}\textbf{Kleene Theorem} (A uniform and weak version of the Second Kleene Theorem)
  There is a script \texttt{uk} such that for all binary scripts \texttt{x}
    we have
  \begin{verbatim}
-> uk x
kx()
  \end{verbatim}
with \texttt{kx} such that, for  all \texttt{z} we have (see \S\ref{bash_notat1}(4) for this equality)
\begin{verbatim}
kx z = x kx z
\end{verbatim} 
In other terms, \texttt{uk} produces uniformly a script \texttt{kx}
such that 
\[
\mathtt{
\varphi_{kx}(z)=\varphi_x(kx,z).
}
\]
So, we can now get the fixed-point uniformly in \texttt{x}.
\ee
\proof We have 
\quad \begin{verbatim}
uk()=
  echo "set k$1 \$1;$(cat $1)">k$1
  chmod 755 k$1
-> uk x
  kx()=
    set kx $1
    x()
    \end{verbatim}
Indeed, when \texttt{x} is assigned to \texttt{\$1}
the line \texttt{echo\ldots} redirects 
 (via a command substitution similar to the one under
part (2) of \S\ref{bash_commands})  the string
\begin{verbatim}
set kx $1;(x)
\end{verbatim}
to file \texttt{kx}. Since the second line of \texttt{uk} makes script \texttt{kx}
executable, we may conclude, by the semantics of \texttt{set},   that \texttt{kx z}
behaves like   \texttt{x kx z}.

\bb\textbf{Example}\quad Let us apply the theorem with \texttt{cat2}
as \texttt{x}  
\quad \begin{verbatim}
-> uk cat2               // uk  with cat2 as x creates executable script kcat2
->  kcat2 id             // kcat2 by input id behaves like cat2 cat2 id
set kcat2 $1;cat $1 $2   // prints (kcat2)
echo $1                  // and (id)
->  cat kcat2            // to check this let's use cat to print directly kcat2
set kcat2 $1;cat $1 $2   // indeed this equals the first output of kcat2
\end{verbatim} 
\ee
\bb\textbf{Quine}\quad By replacing in the example above \texttt{cat2} with
\begin{verbatim}
self()=
  cat $1
\end{verbatim}
we get the rather compact \textit{quine}
\begin{verbatim}
kself()=
  set kself $1;cat $1
-> kself
set kself $1;cat $1
\end{verbatim} 
But of course the quine can bring some extra luggage
\begin{verbatim}
self_plus(:)
  cat $1
  ((  a = 9**9 ))
  echo $a
-> uk self_plus
-> kself_plus
cat $1;((  a = 9**9 ));echo $a
387420489
\end{verbatim}
\ee

 \bb
 \textbf{Definition}\quad A \textit{script-maker}
is a unary script \texttt{x} that for each string \texttt{y}
prints an executable  script
$\mathtt{u=\varphi_x(y)}$
which, in turn, computes a function $\mathtt{\varphi_u(x)}$. That is to say that
for all \texttt{x, y} there is \texttt{u} such that we have
\begin{verbatim}
-> x y
  u()
\end{verbatim} 
and for all \texttt{z} there is \texttt{w} such that
\begin{verbatim}
-> u z
w
\end{verbatim} 
or $\mathtt{\varphi_u}$ is not defined at \texttt{z}.
 \ee
 \bb\label{RogersTh} \textbf{Rogers Theorem}\quad
(A strong and uniform variant of the Second Kleene Theorem) 
There is a script \texttt{ur} that for each  script-maker \texttt{x}
yields  a  script \texttt{krx}
such that, for all \texttt{z}  we have
\[
\mathtt{\varphi_{krx}(z)=\varphi_{\varphi_x(krx)}(z)}
\] 
	 \ee
	 \proof
Define 
\begin{verbatim}
ur()=
  echo "$1 \$1 > ${1}_;chmod 755 ${1}_; ${1}_ \$2"  > r$1
  uk r$1
\end{verbatim}
For \texttt{\$1=x} the two lines of this script produce the two scripts below, one by redirection to \texttt{r\$1=rx}, and the other by application to \texttt{rx} of the uniform procedure of \S\ref{weak_bash}
\begin{verbatim}
-> ur x
  rx()
  krx()
\end{verbatim} 
the form of the former is 
\begin{verbatim}
rx()=
  x $1 > x_
  chmod 755 x_
  x_ $2
\end{verbatim}
and we have
\begin{equation}
\label{rg_bash}
\mathtt{\varphi_{rx}(y,z)=\varphi_{x\_}(z)=\varphi_{\varphi_x(y)}(z)}
\end{equation}
because the script above, when called with arguments \texttt{y, z},
 sends $\mathtt{u=\varphi_x(y)}$ to \texttt{x\_}, then
computes $\mathtt{\varphi_u(z)}$ (line \texttt{x\_ \$2} with \texttt{\$2=z}).

The result now follows by Kleene Theorem since we have
\[
\mathtt{
\varphi_{krx}(z)=\varphi_{rx}(krx,z)
}
\]

\end{document}